# Novel Mössbauer experiment in a rotating system and the extra-energy shift between emission and absorption lines

T. Yarman,[1] A. L. Kholmetskii,[2] M. Arik,[3] B. Akkuş,[4] Y. Öktem,[4] L. A. Susam,[4] O. V. Missevitch[5]

[1] Okan University, Akfirat, Istanbul, Turkey
[2] Department of Physics, Belarusian State University, 4, Nezavisimosti Avenue, 220030, Minsk, Belarus, tel. +375 17 2095482, fax +375 17 2095445, e-mail: khol123@yahoo.com
[3] Bogazici University, Istanbul, Turkey
[4] Istanbul University, Istanbul, Turkey
[5] Institute for Nuclear Problems, Belarusian State University, Minsk, Belarus

**Abstract.** We present the results of a novel Mössbauer experiment in a rotating system, implemented recently in Istanbul University, which yields the coefficient $k=0.69\pm0.02$ within the frame of the expression for the relative energy shift between emission and absorption lines $\Delta E/E=ku^2/c^2$. This result turned out to be in a quantitative agreement with an experiment achieved earlier on the subject matter (A.L. Kholmetskii et al. 2009 *Phys. Scr.* **79** 065007), and once again strongly pointed to the inequality $k>0.5$, revealed originally in (A.L. Kholmetskii et al. 2008 *Phys. Scr.* **77**, 035302 (2008)) via the re-analysis of Kündig's experiment (W. Kündig. *Phys. Rev.* **129**, 2371 (1963)). A possible explanation of the deviation of the coefficient $k$ from the relativistic prediction $k=0.5$ is discussed.

## 1. Introduction

It is known that the first (and major) series of Mössbauer experiments in rotating systems had been carried out at the early 1960's (e.g. [1-6]) soon after the discovery of the Mössbauer effect. In these experiments, an absorber orbited around a source of resonant gamma-radiation (or vice versa). The goal was to verify the relativistic dilation of time for a moving resonant absorber (source), in rotation, which induces the relative energy shift between emission and absorption lines at the value

$$\Delta E/E=\pm ku^2/c^2, \qquad (1)$$

where $k=0.5$ according to relativity theory, $u$ is the tangential velocity of absorber, $c$ is the light velocity in vacuum; the sign "+" corresponds to the case, where a source orbits around an absorber, and the sign "-" replies to the reverse case, where an absorber orbits around a source.

Following Einstein, by then it was indeed anticipated that only the effect of tangential velocity would matter, and the effect of acceleration particularly, would not affect the result [7].

In any case, for sub-sound $u\approx300$ m/s, the ratio $u/c\approx10^{-12}$, so that the energy shift (1) can be reliably measured with iron-57 Mössbauer spectroscopy, which allows us to reach the relative energy resolution $10^{-14}$ and higher.

All of the authors of the mentioned above papers [1-6] reported the value of $k=0.5$ within an accuracy about 1 %, confirming thus the original relativistic prediction with regards to the time dilation effect for a rotating object.

Later the relativistic dilation of time, though with regards to uniform translational motion, had been confirmed with much better precision ($10^{-8}...10^{-9}$) in the experiments on ion beams [8, 9], which deprived physicists of further interest in repetition of Mössbauer experiments in rotating systems.

New wave of interest to the Mössbauer experiments in a rotating system emerged after publication of the paper [10], where serious methodological errors in the available experiments achieved in 1960's were revealed. It is fair to recall that this latter work was stimulated by the predictions made by Yarman, and further developed by him and his colleagues [11, 12].

In the mentioned paper [10] Kholmetskii et al pointed out that almost all of the authors ignored the distortions in the measurement of coefficient $k$ in eq. (1) due to the chaotic mechanical vibrations in the rotor system, which are always present. There was a sole experiment by Kündig [1], which is free of the influence of mechanical vibrations on the measured value of $k$. For this purpose he applied the first order Doppler modulation of the energy of $\gamma$-quanta on a rotor at each fixed rotation frequency $\nu$, implementing a motion of the source along the radius of the rotor towards and backward to the resonant absorber. By such a way, Kündig recorded the shape and the position of resonant line on the energy scale versus the rotation frequency and thus his results were practically insensitive to the presence of possible mechanical vibrations in the rotor system, because such vibrations broaden resonant line, but do not affect its position on the energy scale due to their chaotic nature. In contrast, other authors [2-6] measured only the countrate of detected γ-quanta at a fixed rotational frequency $\nu$, and their results were not protected from the distortions induced by vibrations.

At any rate, via scrupulous analysis of the paper [1], the authors of ref. [10] found a number of computational errors committed by Kündig, and made their own estimation of the coefficient $k$ in eq. (1), based on Kündig's raw data presented in [1]. As a result, they obtained $k=0.596\pm0.006$, which drastically deviates from the relativistic prediction, and exceeds many times (the order of magnitude and more) the reported uncertainty of the experiment [1].

Based on this disclosure, Kholmetskii et al conjectured that in rotating systems, the energy shift between emission and absorption resonant lines is induced not only via the standard time dilation (which is measured alone in the experiments with ion beams [8, 9] dealing with an inertial motion), but also via some additional effect (most likely, due to the direct effect of acceleration, as originally advocated by Yarman et al), thus yielding an excess of $\Delta E/E$ in comparison with the standard relativistic prediction. Our revelation stimulated the performance of our own experiment on the subject matter in 2008 [13], where a novel methodological approach was applied, which, just like in Kündig experiment, allowed eliminating the influence of mechanical vibrations of the rotor on the measured value of coefficient $k$ in eq. (1). As a result, Kholmetskii et al. came up with $k=0.68\pm0.03$ [13], which later has been corrected to the precise values of the Debye temperatures of resonant absorbers [14] as

$k=0.66\pm0.03$. (2)

One should note that the extraordinary result $k>0.5$ is not related to any instrumental error, given that it has been obtained in two different experiments [1] (as re-analyzed in [10]) and [13], which were based on different measurement techniques and data processing procedures. Due to the high fundamental importance of this result, its further experimental verification was strongly required.

In the present paper we thus report the result of measurement of the coefficient $k$ in a recent experiment implemented in Istanbul University (IU), which is similar in its methodology to the previous experiment [13], but uses an enhanced rotor system. In this latter experiment we obtained

$k_{IU}=0.69\pm0.02$, (3)

which manifestly agrees with the result (2).

In section 2 we summarize the present experiment performed in Istanbul University, and the data proceesing procedure. In section 3 we present possible explanations of the extra energy shift between emission and absorption lines, which, to the accuracy $c^{-2}$, is added to the shift caused by the relativistic dilation of time. In sub-section 3.1 we criticize a recent attempt to explain the inequality $k>0.5$ with the hypothesis about the existence of a universal maximal acceleration [15-18], and concurrently show that recent attempts to re-interpret the Mössbauer experiments in rotating systems [18, 19] are, in fact, erroneous. In sub-section 3.2 we present a possible explanation of eqs. (2), (3) on the basis of "conservative" relativity principle [20], which predicts the dependence of time rate of a charged particle on the electric potential at its location. Finally, we conclude in section 4.

## 2. Experiment and data processing

In comparison with the experiment [13], in the novel experiment we applied a rotor with a radius about 2 times smaller (see Table 1, 1st line), which allowed us to increase about 4 times the detector's countrate and to insure a better statistic quality of the data obtained during a given period of measurement time.

In addition, we expanded the range of variation of tangential velocities of absorber (10-260 m/s), which at a smaller rotor radius implies an essential increase of the range of rotational frequencies, thus an increased centrifugal acceleration (see Table 1, 2-4 lines).

Finally, we achieved the air pressure in the rotor chamber less than 1 mmHg (see Table 1, 5 line), where the increase of temperature of rotor due to its heating did not exceed 10°C.[1] This allowed us to neglect any temperature corrections to the Mössbauer spectra of both of the resonant source and resonant absorbers.

In order to detach from each other the contributions of the time dilation effect and mechanical vibrations in the rotor system to the measured countrate of detector at different rotational frequencies, we applied the method, which had been tested and approved throughout the experiment [13].

This method is based on the collection of experimental data for two different resonant absorbers, whose resonant lines are shifted on the energy scale with respect to each other approximately by their linewidth. As we have mentioned above, chaotic vibrations do not affect the position of resonant lines, and cause only their broadening. Hence an equal broadening of shifted lines of these absorbers, caused by vibration, should induce quite different variation of the detector's countrate with the change of rotational frequency.

Therefore, implementing the joint processing of data obtained with both resonant absorbers, we can separate the variation of detector's countrate, caused by the energy shift (1) from the distortions of countrate, caused by the broadening of resonant lines due to vibrations.

The general scheme of our experiment is shown in Fig. 1.

The rotor represents a tube made of ultrastrong aluminium alloy B95 with the mass 64 g. The source [57]Co (Cr) and its lead collimating system, covered by thin cupper and aluminium layers, is located in the middle of the tube, while the absorber of resonant radiation is fixed at the egde of the tube. A balancing mass is placed on the opposite side of the tube. The value of this balancing mass is adjusted on the rotor balance machine, providing the sensitivity about $10^{-3}$ N×m. This corresponds to the uncertainty in the determination of balancing mass less than 1 mg, which, at the indicated rotor mass 64 g, provides equal level of vibrations in the rotor system in the measurement cycles with both resonant absorbers. The rotor is rigidly connected with an asynchronous three-phase motor made by the company Hanning Electro (Austria), bearing a maximum rotation frequency 330 rps. The rotor chamber, connected with a vacuum pump, has a diameter of 350 mm and a height of 300 mm. For safety purpose, the walls of the rotor chamber were made of armor material. The beryllium window for output of resonant gamma-quanta from the rotor chamber has a diameter of 20 mm; the width of beryllium layer is 1 mm. The rotor system has been developed by the company "Praks-M" (Minsk), and it allows a semi-automatic operation with the accuracy of setting the rotational frequency less than 0.1 rps.

The detector, an Ar-Xe proportional counter, is located outside the rotor system; its working window lies in the rotational plane. The detector has 90 % detection efficiency and ≤15 % relative energy resolution with respect to 14.4 keV resonant gamma-quanta.

The diameter of the active part of the source [57]Co (Cr) is 4 mm, the width is 0.1 mm. The source is located in the titanium shell of a cylindrical form with the diameter 6 mm and width

---

[1] The measurements of the rotor temperature were carried out by contact method after each measurement cycle, described below.

6 mm. The activity of the source is 10 mCi, and its active part is adjusted to the axis of rotation with the accuracy ±0.05 mm.

Like in the experiment [13], we carried out measurements with two resonant absorbers $K_4Fe(CN)_6 \times 3H_2O$ (absorber 1), and $Li_3Fe_2(PO_4)_3$ (absorber 2), both enriched by $^{57}Fe$ to 90 %. Each absorber represents a thin film packaged between two beryllium layers of the diameter 19 mm and width 0.5 mm; the surface density of both absorbers is 135 mg/cm$^2$.

The Mössbauer spectra of the absorbers, obtained outside the rotor system with the Mössbauer instrument package MS-2000IP [21] (calibration measurements) are shown in Fig. 2, where we indicate the expected range of variation of the second order energy shift at $k$=0.5 and $k$=1.0. The width of the single resonant line for absorber 1 is equal to 0.290±0.002 mm/s, the isomer shift with respect to zero relative velocity of source and absorber is 0.095±0.001 mm/s. The value of quadrupole splitting of resonant lines of absorber 2 is equal to 0.368±0.002 mm/s, the width of resonant lines for absorber 2 is equal to 0.288±0.003 mm/s the isomer shift of quadrupole doublet with respect to zero relative velocity is 0.570±0.002 mm/s. The left resonant line of absorber 2 is shifted with respect to the resonant line of absorber 1 at (0.290±0.001) mm/s. The value of resonant effect for absorber 1 is 29.0 %, and for absorber 2 it is 22.1 % at a room temperature.

In our measurements we applied the rotational frequencies 10, 160, 185, 200, 220, 240 and 257 rps. At the rotor radius 16.11 cm (see Table 1), the rotational frequency expressed in rps, approximately corresponds to the tangential velocity of resonant absorber in m/s. In particular, the limited rotational frequency 257 rps exactly corresponds to the tangential velocity 260 m/s. Further, we assumed that at the lowest rotational frequency 10 rps, vibrations in the rotor system are still absent[2], so that this value was taken as the reference point, and the numbers of counts of detectors $N(\nu)$ measured at larger rotational frequencies were normalized to the number of counts $N(\nu=10)$.

The measurements were carried out in a cycle mode for both resonant absorbers; each cycle consisted of the measurements of a number of counts of detector of resonant gamma-quanta during 200 s at the rotational frequencies indicated above. In order to prevent heating of electromotor, driving the rotor (happened due to a friction of its bearings), a time break of about 0.5 hour was applied between subsequent cycles. At the activity of the source $^{57}Co$ (Cr) 10 mCi, the average detector's countrate was about 5 pulses/s. Thus at each measurement we accumulated about $10^3$ pulses. We applied 30 cycles for each resonant absorber; the total numbers of counts $N(\nu)$ for each resonant absorber at different rotation frequencies and the corresponding ratios $N(\nu)/N(\nu=10)$ are shown in Table 2. At the mean value of measured counts for both absorbers $\approx 3 \times 10^4$ pulses, the relative statistic error was $1/\sqrt{N} \approx 0.6 \%$.

In the data processing procedure, we considered the coefficient $k$ in eq. (1) and the widths of resonant lines Γ, broadened due to vibrations at different rotation frequencies $\nu$, as the unknown parameters, and the coefficient $k$ does not depend on tangential velocity within the measurement precision. At the first stage, having measured the shapes of resonant lines for both absorbers (Fig. 2), we plotted the expected absorption curves for these absorbers at different rotational frequencies $\nu$ in the idealized case of absence of any vibrations in the rotor system[3]. Such idealized curves are shown in Fig. 3a-b at different hypotheses about the value of 0.5<$k$<1.0. In Fig. 3b we also present the measurement data for absorber 2 (black points), normalized to the number of counts at $\nu$=10 rps, and their deviation from the corresponding idealized curve at the given $k$ allows us to estimate the broadening of resonant line in comparison with its proper width at different $\nu$, as described in ref. [13]. An analytical expression for the chaotic vibrations in the rotor system does not exist; correspondingly, the width of broadened resonant lines as the function of rotation frequency $\nu$ at the adopted value of $k$ (the two-component function $\Gamma_k(\nu)$) can be determined only empirically in suitable discrete points by means of computer simulation according to the algorithm described in ref. [13]. Specifically, at the chosen hypothesis about the value of $k$ (which defines the corresponding idealized absorption curve) and the given rotation frequency $\nu$ (which defines the corresponding point of this curve), we simulate the broadening of resonant line due to vibration, keeping its area and position to be fixed. In such simulation, we simultaneously calculated the change of relative resonant absorption, caused by this broadening of resonant line. When the value of the relative resonant absorption becomes equal to the measured value of this absorption at the given $\nu$, the corresponding width of the line $\Gamma_k(\nu)$ is fixed. In this case, the variation of the given value of $\Gamma_k(\nu)$, corresponding to the variation of detector's countrate within its statistic uncertainly, defines the uncertainty of determination of $\Gamma$ at each particular values of $k$ and $\nu$.

---

[2]Indeed, a level of vibrations in rotor systems is approximately proportional to the rotation frequency in square and thus, it can be practically ignored at $\nu$=10 rev/s in comparison with $\nu \geq$160 rev/s.

[3]In the case, where the calculated energy shift at the given $\nu$ corresponded to some fraction of the channel, we plotted a straight line between nearest neighbouring channels of the experimental curves in Fig. 2. Anyway, the statistic uncertainty in determination of the shapes of resonant lines is practically negligible in comparison with other components of uncertainty of measurement of $k$, indicated below.

As the result, we obtain a set of linewidths $\Gamma_k(\nu)$, which characterize the influence of vibrations on the shape of resonant lines at different $\nu$ and $k$ (see Table 3). As our computer simulation shows, the uncertainty of the ratio $\Gamma(\nu)/\Gamma(\nu=10)$ does not practically depend on $k$ and $\nu$ and is equal $\pm 0.05$.

This result agrees with a rough estimation of the uncertainty in the determination of $\Gamma_k(\nu)$, which can be done in the following way. Consider, for simplicity, a point near the minimum of resonant line. Then, the broadening of this line due to vibrations by 2.5…3.0 times reduces the original value of the resonant effect (about 30 %) to 10…12 %. The relative variation of $\Gamma_k(\nu)$ by 5 % (the value, which we obtained in computer simulation) induces the corresponding variation of the height of resonant line by 5%, too. Hence, the relative variation of measured numbers of counts by detector is about (0.10…0.12)·5 %=0.5…0.6 %, which corresponds to the actual relative uncertainty in measurement of number of counts $1/\sqrt{N} \approx 0.6\%$.

Further, using the two-dimensional function $\Gamma_k(\nu)$ obtained with the absorber 2, at the next stage we determined the corresponding distortions of the idealized curves for absorber 1 due to vibrations, at different values of $\nu$ and $k$. The calculated absorption curves, distorted due to vibrations, are shown in Fig. 4, where the measurement data for absorber 1, normalized to the number of counts at $\nu=10$ rps (black points), are also presented.

The framed algorithm allows us to achieve practically the best fitting of experimental data with two sets of free parameters, i.e. the coefficient $k$ in eq. (1), and the level of vibrations, manifesting as $\Gamma(\nu)$ dependence; the algorithm has been realized with the MathCad Professional software.

Finally, we calculated the standard deviation

$$D(k) = \left[\sum_{\nu}(N(\nu) - N_{rk}(\nu))^2\right]^{1/2}, \quad (4)$$

at different $k$, where the summation is carried out over the values of $\nu=160, 185, 200, 220, 240, 257$ rps; $N(\nu)$ denotes the experimental point (i.e. number of counts at the given $\nu$), and $N_{rk}(\nu)$ stands for the corresponding point of real curve for absorber 1.

Finally we plotted the values of $D$ as the function of $k$, Fig. 5. We see that this function has a sharp minimum at $k=0.69$.

The measurement uncertainty of coefficient $k$ is determined by the following factors:

1. Statistic measurement error of number of counts in Mössbauer spectra of both resonant absorbers (calibration data, see Fig. 2).

3. Error of determination of line width $\Gamma$ at various $\nu$ and $k$, which, in return, is caused by the statistical uncertainty in the measurement of total numbers of counts for absorber 2, when we compare the measurement data for this absorber with the idealized curves (see Fig. 3b).

4. Error in determination of the coefficient $k$ in comparison of real curves and collected data for absorber 1 (see Fig. 4).

Since both Mössbauer spectra presented in Fig. 2 had been obtained with a high statistic quality, the first contribution to the total measurement uncertainly of $k$ is negligible in comparison with factors 2 and 3.

Factor 2 coins the uncertainty in the determination of the width of resonant lines $\Gamma_k(\nu)$, which at the relative statistic error of measurement of $N(\nu)$ of about 0.6 %, is found to be equal ±0.05 mm/s. This leads to the corresponding variation of shapes of real curves for absorber 1, presented in Fig. 4.

In order to model the variation of the shapes of these curves, we applied a procedure developed previously in the implementation of our experiment [13]. Namely, at the obtained value of $k$=0.69, we take six values of $\Gamma_{0.69}(\nu)$ at $\nu$=160, 185, 200, 220, 240 and 260 rev/s, and for each such value we fix 5 sub-values $\Gamma_{0.69}$-$\delta\Gamma$, $\Gamma_{0.69}$-$\delta\Gamma/2$, $\Gamma_{0.69}$, $\Gamma_{0.69}$+$\delta\Gamma/2$, $\Gamma_{0.69}$+$\delta\Gamma$ (where $\delta\Gamma$=0.05), prescribing to them numbers 1, 2, 3, 4, 5, respectively. We thence get six sets made of these numbers. Then, by a random choice, we select one number among {1, 2, 3, 4, 5} for each set to model the variation of linewidth at different rotation frequencies within their measurement uncertainty. For the obtained set of numbers, we recalculated the real curve for absorber 1 and further determined standard deviation between measured and calculated data according to equation (4), and fix the value of $k$, corresponding to the minimum of $D(k)$. Repeating the random choice of numbers {1, 2, 3, 4, 5} for each set, we again recalculate $D(k)$, fix its minimum and implement this procedure for 1000 times. Consequently, we obtain a distribution of the values of $k$, corresponding to minima of the functions $D(k)$ at variable linewidths $\Gamma$ within the uncertainty of their measurement. The half-width of the obtained distribution of values of $k$ happens to be equal to ± 0.02.

Therefore, the final result of our measurement is expressed by eq. (3).

We stress that the method applied for the elimination of influence of chaotic vibrations to a measured value of $k$, suggested for the first time in ref. [13], furnishes an unbiased estimation of $k$, when the resonant line broadening due to vibrations keeps its shape approximately Lorentzian. As Kündig's experiment shows, this is actually the case [1]. At the same time, the uncertainty in the determination of $k$ in our experiment is still 3.5 times larger, than the measurement uncertainty of Kündig experiment, as we reanalyzed (ref. [10]). Such is the cost, which we have paid for the elimination of influence of vibrations via the data processing procedure with two different resonant absorbers (we remind that in the Kündig experiment, the influence of vibrations had been eliminated at the instrumental level).

Further we emphasize that the applied data processing procedure is self-consistent and reversible. This allows us to carry out a cross-check of the obtained result (3), comparing the idealized absorption curves for absorber 1 (Fig. 3a) with the measured absorption curves for this absorber (Fig. 4, black points), in order to get a new set of the values $\Gamma_k(\nu)$; this way the value of $k$ is calculated via the comparison of measured absorption for absorber 2 with real curves for this absorber, corrected to the level of vibrations in the rotor system.

Implementing this procedure, we obtained a new set of linewidths $\Gamma_k(\nu)$, which within the calculation uncertainty practically coincides with the data of Table 3.

Having obtained the values of $\Gamma_k(\nu)$, next we plot the real absorption curves for absorber 2 at different $k$ (see Fig. 6, where the measured data for the second resonant absorber are also shown).

Finally, using the data shown in Fig. 6, we again calculate the standard deviation (4), this time between the experimental points $N(\nu)$ for absorber 2 and points $N_{rk}(\nu)$ for the corresponding real curve of this absorber. The obtained dependence of $D$ on $k$ for the absorber 2 is presented in Fig. 7. Like for the dependence $D(k)$ in Fig. 5, a sharp minimum corresponds to $k$=0.69.

The measurement uncertainty of $k$ was calculated according to the algorithm, described above for absorber 1, and is equal to ± 0.02. Hence, we again arrive at eq. (3).

Thus, like in the previous experiment [13], we have confirmed that $k>0.5$, as anticipated originally by Yarman [11, 12]. It is important to note that the deviation between the measured result and relativistic prediction $k=0.5$ exceeds very many times the measurement uncertainty. Therefore, a plausible physical explanation of this result becomes topical, which we discuss in the next section.

## 3. Proposed explanations for the extra energy shift in Mössbauer rotor experiments

*3.1. Hypothesis about the existence of a universal maximal acceleration in nature*

An attempt to explain the inequality $k>0.5$ was done by Friedman *et al* on the basis of their generalization of special relativity (see [15-18]) with the negation of the clock hypothesis by Einstein (i.e., the independency of clock rate on its acceleration [22]). Thereby the authors postulated the presence of a universal maximal acceleration $a_m$ of nature and proposed a modification of the space-time transformation between uniformly accelerated frames, which then reduces to the usual relativistic transformation in the limit $a_m \to \infty$. In particular, with respect to the second order Doppler effect in rotating systems, they derived an expression (in the case, where the source of radiation is located on the rotational axis) [17]:

$$E = \left(1 + \frac{R\omega^2}{a_m}\right)\left(1 - \frac{R^2\omega^2}{c^2}\right)^{-1/2} E_0,$$

where $E_0$, $E$ are the energies of emitted and absorbed radiation, correspondingly, $R$ is the radial coordinate of absorber, and $\omega$ is the angular rotational frequency. Hence the relative energy shift between emission and absorption lines can accordingly be written as

$$\frac{\Delta E}{E} = \frac{E_0 - E}{E_o} \approx -\frac{R\omega^2}{a_m} - \frac{R^2\omega^2}{2c^2} = -\frac{u^2}{c^2}\left(\frac{1}{2} + \frac{c^2}{Ra_m}\right). \tag{5}$$

Thus, comparing eqs. (1) and (5), one comes up, based on the extended relativity of Friedman *et al*, with

$$k = \frac{1}{2} + \frac{c^2}{Ra_m}. \tag{6}$$

The authors also pointed out that the Mössbauer experiments in rotating systems represent a convenient tool to test their hypothesis due to two reasons:
- high sensitivity of Mössbauer effect to the relative energy shift of resonant lines;
- large centrifugal acceleration achieving in these experiments (up to $10^6$ m/s$^2$), which is directed along the line joining the resonant source and the absorber.

Thanks to these features, the Mössbauer rotor experiments become much more sensitive to the assumed presence of a maximal acceleration $a_m$, than, for example, the experiments in particle physics with any kinds of accelerators. In particular, Friedman *et al* conjectured that the inequality $k>0.5$ can be explained via eq. (6). Taking the result of Kündig experiment, which we have reanalyzed and accordingly corrected in ref. [10] ($k=0.596\pm0.006$), as the most reliable result, they estimated the maximal acceleration as [16]

$$a_m \approx 10^{19} \text{ m/s}^2. \tag{7}$$

This is indeed a huge acceleration from the practical point of view, and it exceeds by many orders of magnitude the typical acceleration of particles in accelerators. However, we notice that the hypothesis about a universal maximal acceleration implies the existence of a fundamental time unit $t_{\text{fundamental}} = c/a_m$, which for the estimated value (7) yields

$t_{\text{fundamental}} \approx 3\cdot10^{-11}$ s.

This time interval corresponds to an electromagnetic radiation with a wavelength of about 1 cm, and this seems to be too large, in order to be considered as the basis of a fundamental time unit.

Rather one can suppose that the fundamental time unit is determined by the relationship

$$t_{fundamental} = l_P/c,$$

where $l_P \approx 1.616 \times 10^{-35}$ m is the Planck length. In this case, the universal maximal acceleration, if it exists, is defined via the equation $a_m = c^2/l_P \approx 5.5 \times 10^{51}$ m/s$^2$. Then eq. (6) reads as $k = \frac{1}{2} + \frac{l_P}{R}$, and the deviation of $k$ from 0.5 becomes non-observable in the Mössbauer rotor experiments.

One more concrete argumentation against the hypothesis by Friedman *et al* is related to the observation that, according to eq. (6), the measured value of $k$ should depend on the rotor radius $R$. Therefore, if eq. (6) along with the estimation (7) is correct, then in the experiment by Kündig [1] (where $R$=9.3 cm) the value of $k$ must be larger than in the experiment by Kholmetskii *et al* [13] ($R$=30.5 cm). However, this is not the case.

This observation is once again strongly against the hypothesis by Friedman *et al*, at least with the value of maximal acceleration (7).

Being not satisfied with this outcome, Friedman & Nowik [18] claimed that only the experiment by Kündig is the correct one (because, as we mentioned above, he was the only one, who measured the shape of resonant lines at each rotational frequency), whereas, according to Friedman & Nowik, the results of all other Mössbauer rotor experiments, including Kholmetskii *et al* experiment [13], are erroneous.

In order to substantiate this assertion, Friedman & Nowik carried out in [18] their own calculation of the relative energy shift between emission and absorption lines for the configuration, where a resonant absorber rotates, while a source of resonant radiation is *at rest* in a laboratory frame, and thus do not spin on the rotor axis. For this configuration, a strong aberration effect does emerge, which leads to the substantial broadening of resonant line as a function of finite sizes of source, absorber and the divergence of gamma-beam.

Based on this result, Friedman & Nowik claimed that the broadening of resonant line, observed directly in the experiment by Kündig, takes place due to this aberration effect, but not due to mechanical vibrations in the rotor system, on the contrary to what Kündig assumed. Furthermore, Friedman & Nowik asserted that the experiment by Kholmetskii *et al* [13] as well as all other Mössbauer experiments in rotating systems [2-6] are all erroneous, due to the missed aberration effect, they introduced in ref. [18].

However, the calculations by Friedman & Nowik had been carried out for the configuration, which was not realized in the performed Mössbauer experiments in rotating systems [1-6, 13], where both the source and absorber were *rigidly* fixed on a rotor, and even if the source is located on the rotational axis, it is still to be considered along with the rotational motion. In other words, in the configuration, realized in the Mössbauer experiments [1-6, 13] (and in the present one as well), the source is *at rest in the rotational frame*, but not in the laboratory frame.

Let us show that for this configuration, the aberration effect, calculated by Friedman & Nowik and causing the component of relative energy shift proportional to ($u/c$), completely disappears, and the entire energy shift between emission and absorption lines is proportional to the ratio $u^2/c^2$ (to the accuracy of calculations $(u/c)^2$), regardless of sizes of source, absorber, and divergence of gamma-beam.

In order to prove this statement, it is sufficient to show that for two arbitrary points A and B on the rotor surface (see Fig. 8, where A stands for the point-like source, and B for the point-like absorber), the relative energy shift between emission and absorption lines does not contain the linear terms of order ($u/c$).

For a laboratory observer, the frequency of emitted gamma-quanta is equal to [23]

$$\nu_{em} = \frac{\nu_0 \sqrt{1 - u_A^2/c^2}}{\left(1 - \frac{\mathbf{n} \cdot \mathbf{u}_A}{c}\right)},$$

where $\nu_0$ in the proper frequency of gamma-quanta, $\boldsymbol{u}_A$ is the velocity of point A at the emission moment, and $\boldsymbol{n}$ is the unit vector along the line, joining point A at the emission time moment with point B at the absorption time moment.

Correspondingly, the frequency of absorbed radiation reads as

$$\nu_{ab} = \frac{\nu_{em}\left(1 - \frac{\boldsymbol{n}\cdot\boldsymbol{u}_B}{c}\right)}{\sqrt{1 - u_B^2/c^2}} = \frac{\nu_0\sqrt{1 - u_A^2/c^2}\left(1 - \frac{\boldsymbol{n}\cdot\boldsymbol{u}_B}{c}\right)}{\sqrt{1 - u_B^2/c^2}\left(1 - \frac{\boldsymbol{n}\cdot\boldsymbol{u}_A}{c}\right)}, \qquad (8)$$

where $\boldsymbol{u}_B$ is the velocity of point B at the absorption moment.

In order to calculate the frequency (8), we designate $r_A$, $\vartheta_A$ the radial and angular coordinate of the point A at the moment of emission of gamma-quantum, and $r_B$, $\vartheta_B$ the radial and angular coordinates of the point B at the moment of absorption of gamma-quantum, correspondingly, see Fig. 8. With these designations, we have the following components:

$$n_x = \frac{(r_{AB})_x}{r_{AB}} = \frac{r_B\cos\vartheta_B - r_A\cos\vartheta_A}{r_{AB}}, \quad n_y = \frac{(r_{AB})_y}{r_{AB}} = \frac{r_B\sin\vartheta_B - r_A\sin\vartheta_A}{r_{AB}}, \qquad (9\text{a-b})$$

$$u_{Bx} = \omega r_B\sin\vartheta_B,\ u_{By} = -\omega r_B\cos\vartheta_B,\ u_{Ax} = \omega r_A\sin\vartheta_A,\ u_{Ay} = -\omega r_A\cos\vartheta_A, \qquad (10\text{a-d})$$

where $r_{AB}$ is the distance between the point A at the emission time moment and point B and absorption time moment.

Hence, substituting eqs. (9) and (10) into eq. (8), we derive:

$$\nu_{ab} = \frac{\nu_0\sqrt{1 - u_A^2/c^2}\left(1 - \frac{\boldsymbol{n}\cdot\boldsymbol{u}_B}{c}\right)}{\sqrt{1 - u_B^2/c^2}\left(1 - \frac{\boldsymbol{n}\cdot\boldsymbol{u}_A}{c}\right)} = \frac{\nu_0\sqrt{1 - u_A^2/c^2}\left(1 - \frac{n_x u_{Bx} + n_y u_{By}}{c}\right)}{\sqrt{1 - u_B^2/c^2}\left(1 - \frac{n_x u_{Ax} + n_y u_{Ay}}{c}\right)} =$$

$$= \frac{\nu_0\sqrt{1 - u_A^2/c^2}\left(1 - \frac{(r_B\cos\vartheta_B - r_A\cos\vartheta_A)\omega r_B\sin\vartheta_B - (r_B\sin\vartheta_B - r_A\sin\vartheta_A)\omega r_B\cos\vartheta_B}{r_{AB}c}\right)}{\sqrt{1 - u_B^2/c^2}\left(1 - \frac{(r_B\cos\vartheta_B - r_A\cos\vartheta_A)\omega r_A\sin\vartheta_A - (r_B\sin\vartheta_B - r_A\sin\vartheta_A)\omega r_A\cos\vartheta_A}{r_{AB}c}\right)} =$$

$$\frac{\nu_0\sqrt{1 - u_A^2/c^2}}{\sqrt{1 - u_B^2/c^2}}. \qquad (11)$$

Thus, we see that the terms of nominator and denominator, which contain the linear components in $(u/c)$, mutually cancel each other, so that the frequency (energy) shift is determined by the second order Doppler shift alone, if we do not include the extra energy shift, discussed in the present paper.

Since equation (11) was derived for two arbitrary points A and B on a rotor surface, it also remains in force for any spatially extended source and absorber, and does not depend on the divergence of the gamma-beam. The only point is that for such spatially extended source centered on the rotational axis, the tangential velocities $u_A$ at the edge of the source and at its center, differ from each other, which can cause the broadening of the emitting resonance line. However, for a source of resonant radiation, sufficiently compact and bearing typical configurations of Mössbauer rotor experiments, this effect is quite negligible. For example, in the present experiment, with the diameter of source $^{57}$Co 4 mm, we have $r_A$=2.0 mm on its periphery. For the maximal rotation frequency $\nu$=260 rev/s, the corresponding velocity is equal to $u_A=2\pi\nu r_A \approx$ 3.3 m/s, and the ratio $u_A^2/c^2 \approx 1.2\times 10^{-16}$. This ratio is well below the sensitivity of iron Mössbauer spectroscopy to the relative energy shifts of resonant lines ($10^{-13}...10^{-15}$) and can be completely ignored. Thus for the practical purpose we can well put $u_A$=0 for a compact source, so that the relative frequency (or energy) shift becomes:

$$\frac{\Delta E}{E} = \frac{v_0 - v_{ab}}{v_0} = 1 - \frac{1}{\sqrt{1 - u_B^2/c^2}} \approx -\frac{u_B^2}{2c^2},$$

when the extra energy shift is not included.

This result shows the irrelevance of the analysis by Friedman & Nowik [18] implemented for the case, where the source rests in the laboratory frame. For real configurations of all of the Mössbauer experiments in a rotating system, where both the source and the absorber are rigidly fixed on a rotor, eq. (1) remains valid, and confirms the correctness of the methodological approach of these experiments. Therefore, the result of Kholmetskii *et al* experiment [13], just like the result of Kündig experiment [1], appears to be correct and, taken together, does in effect, disprove the hypothesis by Friedman *et al* about the existence of a universal maximal acceleration in nature, at least at its speculated numerical value (7).

We can add that one more recent attempt by Zanchini [19] to re-interpret the Mössbauer experiments in rotating systems, where a possible contribution of the first order Doppler shift to the measured effect is discussed, is also erroneous, as explained in ref. [24].

Thus, we have to conclude that in the Mössbauer experiments in rotating systems, the energy shift between emission and absorption lines is fully determined by the time dilation effect between a source of resonant radiation and resonant absorber. Therefore, the result $k>0.5$ unambiguously indicates that the tame rate for the nuclei of orbiting absorber experiences an extra dilation (in addition to the classically presumed relativistic effect), and thus, we have to seek an origin of this effect.

*3.2 "Conservative" relativity principle and its experimental verification in Mössbauer experiments in rotating systems*

In our recent paper [20] Kholmetskii et al advanced a new relativity principle, we have named the "conservative relativity principle" (CRP). It represents a generalization of special relativity principle by Einstein to the case of combined action of fields of different nature (e.g., electromagnetic and gravitational) on a given particle, and it is postulated in the following form:

- it is impossible to distinguish the state of rest of any system and the state of its motion with constant velocity, if this system receives *no work* during its motion.

In our opinion, this principle looks very attractive (and, we believe, natural), though, as any other postulate, it must anyway be verified experimentally.

In this respect we stress the principal implication of CRP; it is the dependence of the time rate of a charged particle on the electric potential $\varphi$ at its location. Namely, for a particle with the rest mass *m* and electric charge *e*, this dependence is expressed by the equation [20]

$$d\tau_e = \left(1 + e\varphi/mc^2\right)\frac{dt}{\gamma}, \qquad (12)$$

where *dt* is the time interval, measured in empty space for a resting charged particle outside the electric field, and $\gamma$ is the Lorentz factor. One has to recall that such an occurrence was originally predicted by Yarman [25], with regards to the bound muon decay rate retardation, which afterwards was well checked out in ref. [26].

Eq. (12) can be subjected to experimental test in the analysis of numerous data of precise physics of simple atoms. As shown in [20], in all cases, where the results of QED calculations and experimental data are already in agreement with each other, the corrections to the atomic energy level, induced by CPR, either disappear, or exhibit values less than the measurement uncertainty. However, the available long-standing discrepancies between QED calculations and experiments (e.g., for 1S-2S interval in positronium and 1S hyperfine splitting in positronium) are fully eliminated via the consideration of the requirements imposed by CRP. We also should like to mention that eq. (12) yields the same estimation (though with different uncertainties) for the proton size in the classic 2*S*-2*P* Lamb shift in hydrogen, 1*S* Lamb shift in hydrogen, and 2*S*-2*P* Lamb shift in muonic hydrogen, with the mean value $r_E$=0.841 fm [27], which is in a perfect agreement with the latest experimental data [28]. Finally, eq. (12), as stated, straightforwardly eliminates the deviation between theory and experiment in the measurement of decay rate of bound muons in meso-atoms versus their atomic numbers $Z$ [20, 26].

With respect to Mössbauer experiments in rotating systems, we are inclined to assert that eq. (12) may provide a key regarding the understanding of the origin of the second order extra energy shift, originally revealed through the analysis by Kholmetskii et al of the experiment by Kündig [10], then our own subsequent previous experiment [13], and furthermore, our recent own experiment reported here. The specific feature of such experiments is the huge centrifugal acceleration of resonant absorber, located on the rotor rim (of about $10^5$-$10^6$ m$^2$/s), which is never achieved in any other experiments with condensed matter, both in laboratory and cosmological scales. Therefore, the resonant nuclei of absorber are expected to experience the effect of a local electric field, that would be created to counteract the centrifugal force with the appearance of corresponding local electric potential on the nuclei, never achievable in usual conditions (i.e., outside the rotor system).

Our calculations based on CRP, with regards to the additional second order energy shift for resonant nuclei of absorber are given in ref. [20], and they yield the approximate value $k \approx 0.7$ in a good agreement with our measurements.

Finally, paying attention on some deviation between Kündig result $k=0.596 \pm 0.006$ as we re-estimated [10] and the result of the present experiment (3), we point out that Kündig did not estimate possible variation of characteristics of oscillating piezo-element driving the source [1] due to centrifugal forces, which could lead to a systematic error of his measurements. In contrast, we like to emphasize that our own estimate (3) appears to be unbiased.

## 4. Conclusion

We emphasize that the observed substantial deviation of the coefficient $k$ in eq. (1) from the relativistic value $k=0.5$ in Mössbauer experiments in rotating systems does not yet bear any explanation in the framework of the standard approach. At the same time, it is highly unbelievable that the experimental results of [1] (as re-analyzed in [10]), [13], and the present experiment as well, all giving the substantial excess of $k$ over the relativistic value 0.5, indicate any violation of the well-developed and recognized special theory of relativity. Rather we assume the presence of some missing points in relativistic physics; one of them could be the "conservative" relativity principle, representing the generalization of special relativity principle, and postulated in the form easily adopted by plain intuition. A principal implication of this novel relativity principle is the dependence of time rate of a charged particle on the electric potential at its location. As we have shown in ref. [20], this effect can play an essential role in the Mössbauer experiments in rotating systems, and finally explains the observed inequality $k>0.5$.

## Acknowledgment

The present work was achieved in the framework of the Research Project BAP 5623 of Istanbul University. Thus, we thank Prof. Yunus Söylet, Prof. S. Yarman, Prof. L. Kavut, Prof. A. Akan, Mrs. S. K. Aydın, Mr. O.R. Küçükömeroğlu, and Ms. Heval Kara, who with their positive approach made possible the realization of it. We further thank Dr. Ali Altintas, Dr. Fatih Ozaydin, Dr. Sinan Bugu, Dr. Ozan Yarman, our research assistants Ozgur Aytan, Can Yesilyurt, Gulfem Susoy Dogan and Aysegul Ertoprak, as well as our students, now our colleagues, Agageldi Muhammetgulyyew, Sefika Ozturk Cokcoskun, Duygu Tunçman, Ecem Çevik and Gulce Özmansur, for their valuable efforts regarding the realization of the experiment achieved in Istanbul University.

We thank at the same time, Dr. B. Rogozev and Dr. M. Silin (Ritverc Company, S.-Petersburg, Russia) for preparation and test of both resonant absorbers.

# References

1. W. Kündig. Phys. Rev. **129**, 2371 (1963).
2. H.J. Hay et al. Phys. Rev. Lett. **4**, 165 (1960).
3. H.J. Hay. In: Proc. of Second Conference on the Mössbauer effect, ed. by A. Schoen and D.M.T Compton. Wiley, New York. 1963, p. 225.
4. T.E. Granshaw and H J. Hay. In: Proc. Int. School of Physics, "Enrico Fermi". Academic Press, New York. 1963, p. 220.
5. D.C. Champeney and P.B. Moon. Proc. Phys. Soc. **77**, 350 (1961).
6. D.C. Champeney, G.R. Isaak and A.M. Khan. Proc. Phys. Soc. **85**, 83 (1965).
7. A. Einstein. The Meaning of Relativity. Princeton University Press, Princeton. 1953.
8. R.W. McGowan et al. Phys. Rev. Lett. **70**, 251 (1993).
9. I. Bailey et al. Nature **268**, 301 (1977).
10. A.L. Kholmetskii, T. Yarman and O.V. Missevitch. Phys. Scr. **77**, 035302 (2008).
11. T. Yarman. Found. Phys. Lett. **19**, 675 (2006).
12. T. Yarman, V.B. Rozanov and M. Arik. In: Proc. Int. Conf. Physical Interpretation of Relativity Theory. Moscow Bauman State University, Moscow. 2007, p. 187-197.
13. A.L. Kholmetskii, T. Yarman, O.V. Missevitch and B.I. Rogozev. Phys. Scr. **79**, 065007 (2009).
14. A.L. Kholmetskii, T. Yarman and O.V. Missevitch. Int. J. Phys. Sci. **6**, 84 (2011).
15. Y. Friedman and M. Semon. Phys. Rev. E **72**, 026603 (2005).
16. Y. Friedman and Yu. Gofman. Phys. Scr. **82**, 015004 (2010).
17. Y. Friedman. Ann. Phys. **523**, 408 (2011).
18. Y. Friedman and I. Nowik. Phys. Scr. **85**, 065702 (2012).
19. E. Zanchini. Phys. Scr. **86**, 015004 (2012).
20. A.L. Kholmetskii, T. Yarman and O.V. Missevitch. Eur. Phys. J. Plus **129**, 102 (2014).
21. A.L. Kholmetskii, V.A. Evdokimov, M. Mashlan et al. Hyperfine Int. **156/157**, 3 (2004).
22. A. Einstein. Ann. Phys. Lpz. **35**, 898 (1911).
23. C. Møller. The Theory of Relativity. Clarendon Press, Oxford. 1973.
24. A.L. Kholmetskii, O.V. Missevitch and T. Yarman. Phys. Scr. **89**, 067003 (2014).
25. T. Yarman. In: DAMOP 2001 Meeting, Session J3. Book of abstracts. APS, London, Ontario. 2001.
26. T. Yarman, A. L. Kholmetskii and O. V. Missevitch. Int. J. Theor. Phys. **50**, 1407-1416 (2011).
27. A.L. Kholmetskii, O.V. Missevitch and T. Yarman. Can. J. Phys. **92**, 321 (2014).
28. A. Antognini, F. Nez, F. Schuhmann et al. Science **339**, 417 (2013).

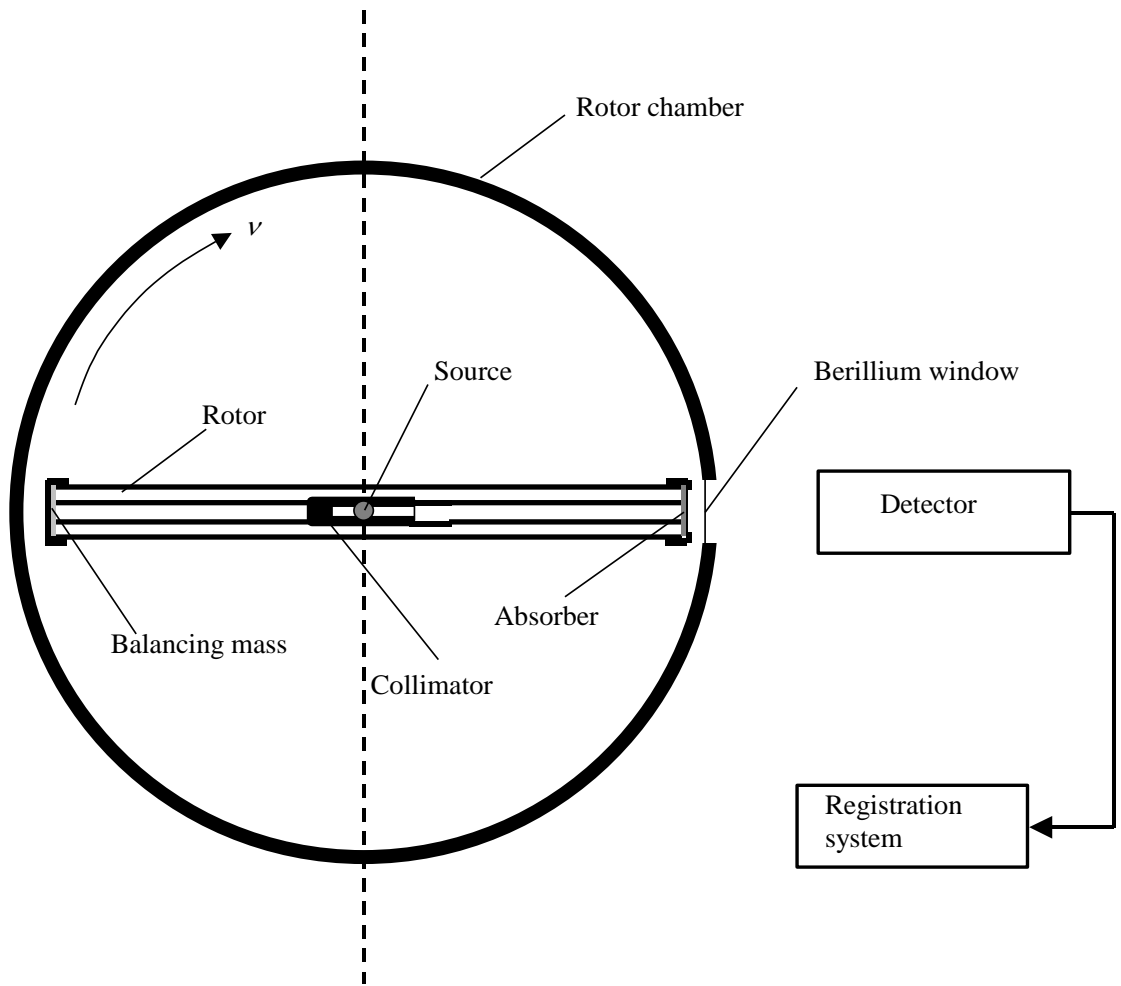

Fig. 1. General scheme of Mössbauer experiments in rotating systems. A source of resonant radiation is located on the rotational axis; an absorber is located on the rotor rim, while a detector of gamma-quanta is placed outside the rotor system, and it counts gamma-quanta at the time moment, when source, absorber and detector are aligned in a straight line.

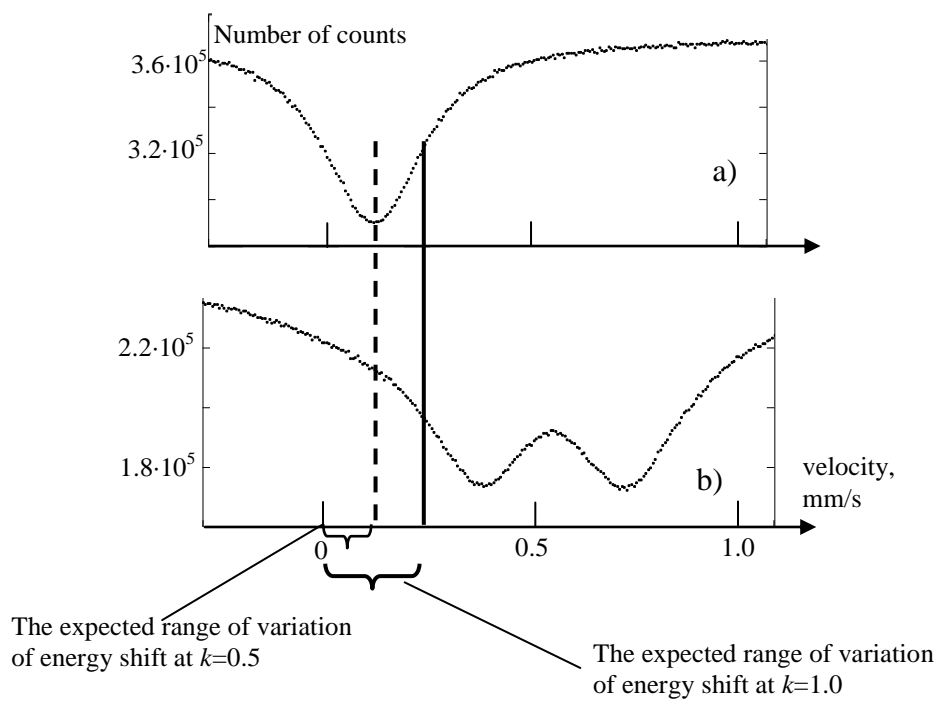

Fig. 2. Mössbauer spectra of absorber 1 $K_4{}^{57}Fe(CN)_6 \times 3H_2O$ (a) and absorber 2 $Li_3{}^{57}Fe2(PO_4)_3$ (b), obtained with the source $^{57}Co(Cr)$ at room temperature $T_R=295\pm2$ K (calibration data), and the expected range of variation of energy shift between emission and absorption lines in our rotor experiment for two limited hypotheses about $k$.

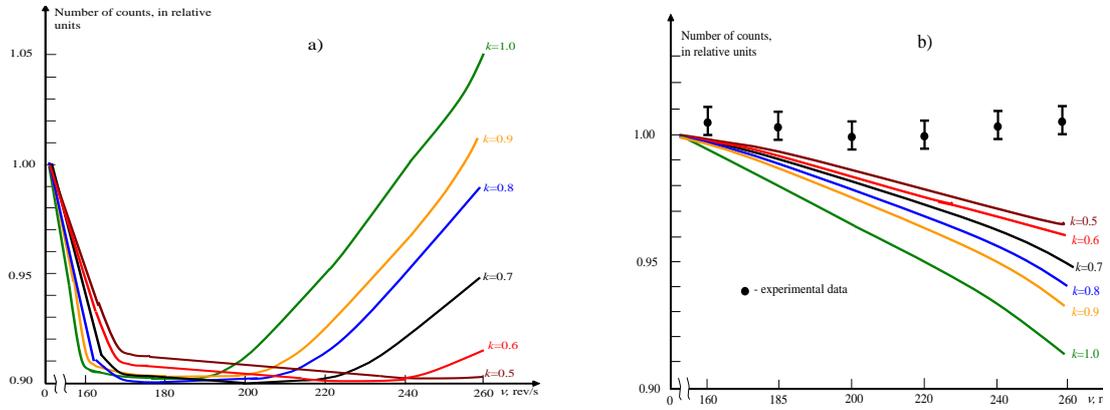

Fig. 3. Idealized absorption curves for absorber 1 (a) and absorber 2 (b) calculated via eq. (1) for the shapes of resonant lines of both absorbers presented in Fig. 2 (no vibration case). Black points show the experimental data obtained with absorber 2, plotted in the relative units, where the error bar has a purely statistic origin.

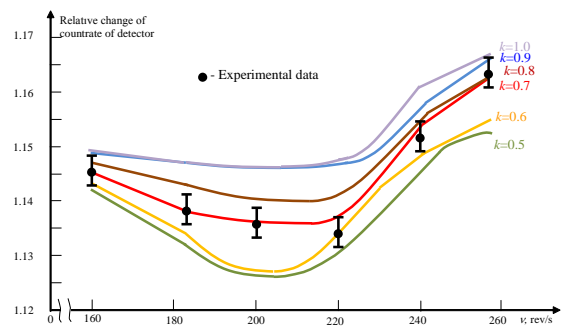

Fig. 4. Real absorption curves for absorber 1 at different *k* in comparison with the experimental data, plotted in the relative units (black points, where the error bar has a purely statistic origin).

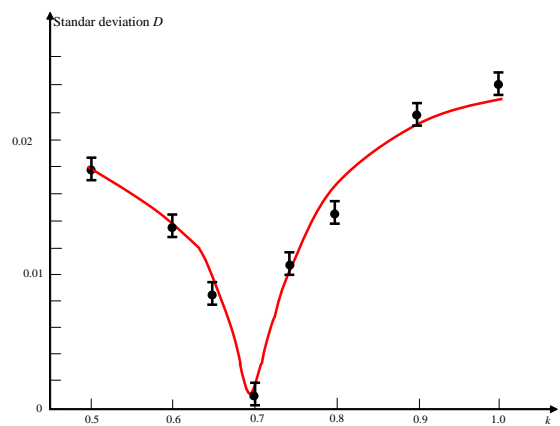

Fig. 5. Standard deviation between measured and calculated data for absorber 1 versus *k*.

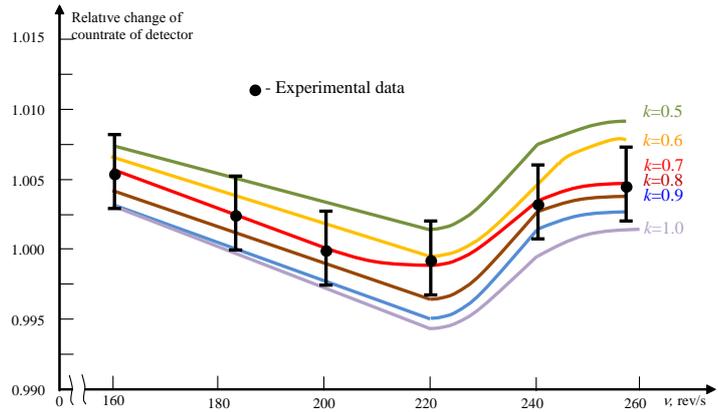

Fig. 6. Real absorption curves for absorber 2 at different *k* in comparison with the experimental data for this absorber (black points).

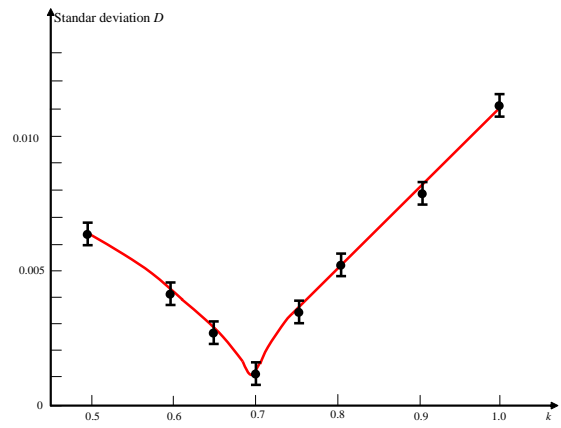

Fig. 7. Standard deviation between measured and calculated data for absorber 2 versus $k$.

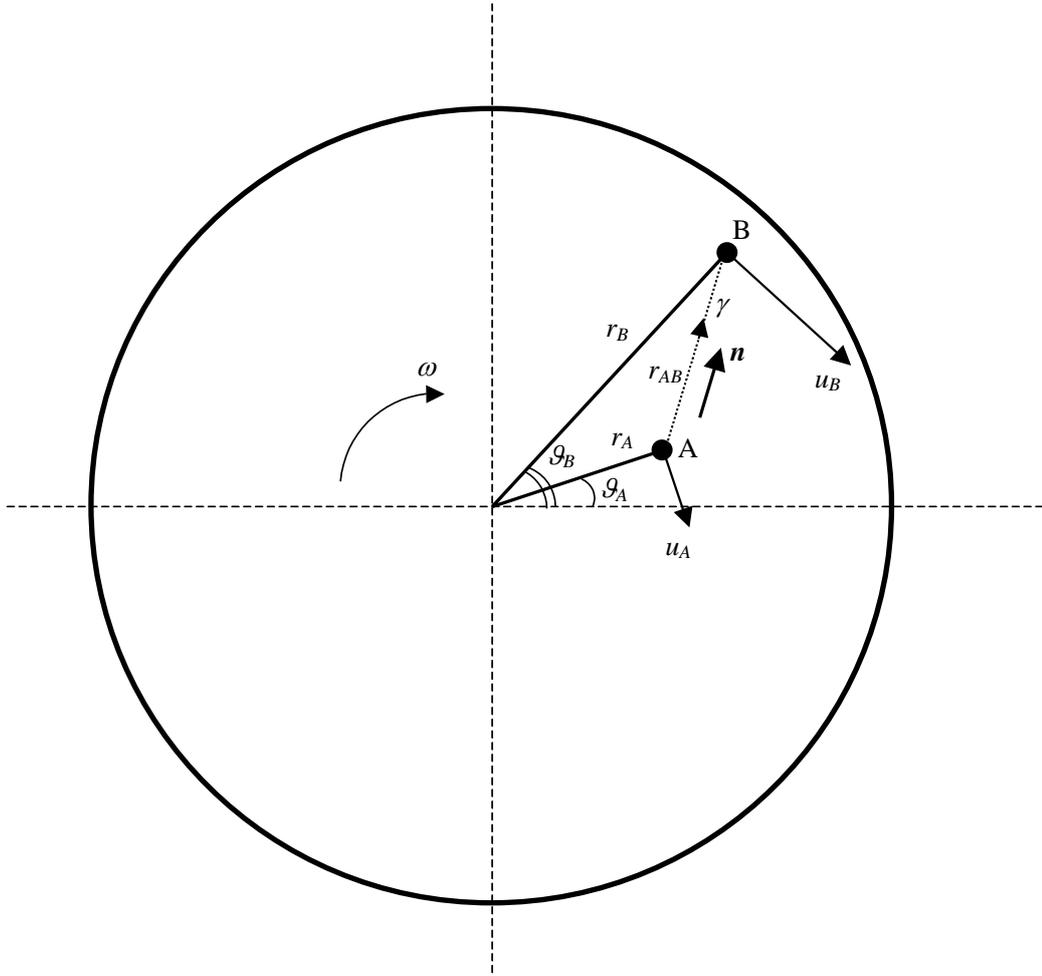

Fig. 8. Diagram for calculation of the Doppler effect in a rotating system between a point-like emitter (located in the point A at the emittance moment) and point-like receiver (located in the point B at the receiving moment).

Table. Comparative characteristics of the rotors systems in the present experiment and in experiment [13]

|  | Present experiment | Experiment [13] |
|---|---|---|
| Rotor radius, cm | 16.11 | 30.10 |
| The range of rotational frequencies, rev/s | 10-260 | 70-110 |
| Velocity range of absorber, m/s | 10-260 | 140-220 |
| Maximal centrifugal acceleration, m/s$^2$ | $4.2 \times 10^5$ | $1.4 \times 10^5$ |
| Pressure in the rotor chamber, mmHg | < 1 | < 50 |

Table 2. Measured number of counts $N(\nu)$ or absorbers 1 and 2 at different rotational frequencies $\nu$ and corresponding relative values $N(\nu)/N(\nu=10)$

| Rotation frequency | | 10 | 160 | 185 | 200 | 220 | 240 | 260 |
|---|---|---|---|---|---|---|---|---|
| Absorber 1 | $N(\nu)$ | 30865 | 35379 | 35128 | 35065 | 35010 | 35558 | 35931 |
| | $N(\nu)/N(\nu=10)$ | 1.000 | 1.146 | 1.138 | 1.136 | 1.134 | 1.152 | 1.164 |
| Absorber 2 | $N(\nu)$ | 30551 | 30732 | 30644 | 30547 | 30562 | 30688 | 30710 |
| | $N(\nu)/N(\nu=10)$ | 1.000 | 1.006 | 1.003 | 1.000 | 1.000 | 1.004 | 1.005 |

Table 3. Broadening of resonant line for absorber 2 expressed as the ratio $\Gamma_r=\Gamma(\nu)/\Gamma(\nu=10)$ at different rotational frequencies $\nu$ and different coefficients $k$. The uncertainty in the determination of ratio $\Gamma(\nu)/\Gamma(\nu=10)$ is equal to ±0.05

| $k=0.5$ | | $k=0.6$ | | $k=0.7$ | | $k=0.8$ | | $k=0.9$ | | $k=1.0$ | |
|---|---|---|---|---|---|---|---|---|---|---|---|
| $\nu$ | $\Gamma_r$ | $\nu$ | $\Gamma_r$ | $\nu$ | $\Gamma_r$ | $\nu$ | $\Gamma_r$ | $\nu$ | $\Gamma_r$ | $\nu$ | $\Gamma_r$ |
| 10 | 1.0 | 10 | 1.0 | 10 | 1.0 | 10 | 1.0 | 10 | 1.0 | 10 | 1.0 |
| 160 | 2.85 | 160 | 2.90 | 160 | 2.95 | 160 | 3.05 | 160 | 3.05 | 160 | 3.10 |
| 185 | 2.75 | 185 | 2.75 | 185 | 2.80 | 185 | 2.95 | 185 | 3.05 | 185 | 3.05 |
| 200 | 2.55 | 200 | 2.60 | 200 | 2.80 | 200 | 2.90 | 200 | 3.05 | 200 | 3.00 |
| 220 | 2.70 | 220 | 2.75 | 220 | 2.80 | 220 | 2.90 | 220 | 3.05 | 220 | 3.05 |
| 240 | 3.00 | 240 | 3.10 | 240 | 3.20 | 240 | 3.20 | 240 | 3.40 | 240 | 3.45 |
| 257 | 3.20 | 257 | 3.25 | 257 | 3.45 | 257 | 3.45 | 257 | 3.45 | 257 | 3.50 |